# On the Aharonov-Bohm Effect
# and Why Heisenberg Captures Nonlocality Better Than Schrödinger

Yakir Aharonov

*School of Physics and Astronomy, Tel-Aviv University,*
*Ramat-Aviv 69978, Israel,*
*Institute of Quantum Studies and Faculty of Physics, Chapman University,*
*1 University Drive, Orange, CA 92866, USA, and*
*Iyar, The Israeli Institute for Advanced Research,*
*Rehovot, Israel*
*E-mail: yakir@post.tau.il, aharonov@chapman.edu*

I discuss in detail the history of the Aharonov-Bohm effect in Bristol and my encounters with Akira Tonomura later on. I then propose an idea that developed following the publication of the Aharonov-Bohm effect, namely the importance of modulo momentum and Heisenberg representation in dealing with non-local quantum phenomena.

## 1. Introduction

It is both a sad and a happy occasion for me. Sad because Professor Akira Tonomura is not with us anymore. He was a good friend of mine. We first met about 30 years ago in a conference that he organized in Hitachi. It was a very interesting conference and also an interesting meeting for me with him. He was very interested in my work on the Aharonov–Bohm (A–B) effect[1] and other questions of quantum mechanics. Since then we met many times and I always found him extremely enthusiastic about physics, with a lot of knowledge, not only about experimental physics, but also on theoretical questions. I must say that the first time he told me that he was going to try to do the A-B effect with a completely closed flux inside the superconductor tube, I did not want to tell him that it is possible, but I was sure that it would not work, because you needed to create a very, very small tube and to make sure that the beam of the electrons around it is still coherent. But he *was* successful and eventually performed an amazing experiment! This, then, is the happy side of this occasion. I think this experiment is one of the most beautiful ones in modern physics. I was hoping that he would be able to complete this very ambitious attempt to build the best electron microscope in the world. I now hope that his team will continue to work on it in his memory and complete his life goal.

## 2.  Historical recollection of development of the Aharonov-Bohm effect

Let me begin by telling you a little about the history of the A-B effect, how I came to think about it, and how I collaborated with David Bohm, who was my thesis advisor at that time at Bristol University. It started, in effect, because I was a graduate student and a very ignorant one.  Being ignorant, I knew nothing about gauge transformations. I did not know that you are not supposed to think about a potential, which is only a function of time, as something that should not have any effect at all. If I knew that, I wouldn't find the effect.

I was very excited when I learned about the Bloch functions. These are quantum functions. When you have a periodic potential in position and you look at the energy level, you find that there is a gap in energy because of the degeneracy for positive and negative momentum. If the periodic potential connects them, you get a gap between the two energies that are separated by this potential. So, as an ignorant student I thought the following: Why not look for an analogy? If I exchange position with time, and momentum with energy, then I'd have a periodic potential, which is a function of time. I will have gaps, not in energy, but in momentum. So, I tried to play around. I knew how to solve Schrödinger's equation. Then I added the periodic time-dependent potential to the equation. Lo and behold! The only thing I got, as everybody knows today, but *I* didn't know at that time, is just a phase. I kept on getting a phase, and a phase doesn't do anything, because you can't measure it! I was very frustrated. That night I could not sleep, and in the morning it occurred to me that maybe if I have, in one region of space, one potential which is only a function of time, I will get a phase, and in another region of space, where I have a different potential that is only a function of time, I will get a *different* phase. And then – maybe – I will see the relative phase as having a meaning.

So I thought about looking at two Faraday cages. In one cage, I have half of the electrons and the other half of the electrons in the other cage. I next considered the superposition of the electrons in both of them. In one Faraday cage I put the potential. So, inside it there will be nothing, just the potential. In the other Faraday cage, I will ground it so there will be no potential. And then when you solve the Schrödinger equation, you find indeed that in the first Faraday cage you do get the phase, which is the integral of the potential of the time proportional to it. In the other cage you get no potential. Now you remove the potential, so there is no effective field anywhere. Luckily, the phase still "remembers" not what the potential is now, but what the integral of the potential was. So now I open the two cages, send the two parts of the beam together. Lo and behold! You get an effect, which is a change in interference, even though the particles never touched the electric field.

So I came to my professor, David Bohm, and I showed it to him. And he said, "That's interesting. That's interesting. It's an interesting idea, but we have to think some more what we can do with it." And it lay there for a few months. And we thought about it only as a curiosity, nothing more. We did not connect it yet with the basic idea of electromagnetic potential.

Then Bohm sent me to a summer school that was at Oxford. At that time, it was very popular for people to think about physics in the language of analyticity. This was the time when people did not know yet how to renormalize field theories. I confess it was very boring.

Anyhow I was listening to some talks there, and there was one talk about what happens when you perform scattering with a vector potential. Suddenly it struck me that maybe there could also be an effect similar to the scalar potential with the vector potential. Excited I came back to Bristol and told Bohm, "Look what happens if, instead of just two different time-dependent scalar potentials, we have a solenoid with a vector potential around it. We should get an effect analogous to the one that we discussed before." Then *he* got also excited and said, "Okay, first of all, let's modify it. Instead of your two Faraday cages let's talk about two long cavities when you have different potentials. We send the two wave packets of the electron through them, and when the wave packets are inside the cavities we switch the potential in one of them and switch it off before the wave packet gets out." Then he said, "In order to really be able to publish it, I would like you to solve the problem of what happens if you have an infinitely thin solenoid and you do a scattering of electron beams when they go through this infinitely thin solenoid. Classically, they should not be affected at all, but quantum mechanically there should be some interesting scattering.

So I had to solve that problem. And it turned out that in order to solve it you have to sum up infinite series of Bessel functions, but unfortunately, with fractional order. So if you have to sum up Bessel functions, you only look at the Watson mathematical encyclopedia. I looked at the Watson, and there was nothing like that. I was walking around, didn't know what to do. Luckily, there was the chairman of the Physics Department at Bristol University, Price. And when Price looked at me and saw me walking like this, he asked, "What's the problem?" I told him, "I don't know how to sum these series." He said, "I think that there could be some interesting differential equation that could be connected to this. And if you could solve this, you will be able to continue."

I took his suggestion, and indeed was able to do it, and solved the exact formula for the scattering. And, in fact, Bohm and I suggested to Price that he would also be a coauthor of this article. But he decided that he did not contribute enough to deserve that.

In Bristol, like in any other universities in England, there is a tradition of afternoon tea. Professors and graduate students get together, discussing different problems in physics. In one of these meetings, I told them that there is this interesting new possibility of seeing an effect on charges without field. But I said, "Unfortunately, I don't believe that it could be observed experimentally because we will need to have an extremely thin solenoid in order to have the possibility of seeing the effect." Among the professors attending that afternoon tea was Sir Frank, who was an expert on solid state, and he said, "Hey, hey, I believe it can be done with magnetic whiskers." There are magnetic whiskers which are very, very thin, line of magnetic flux inside crystals. And then Chambers was sitting next to him, and he said, "Okay, I'm going to do the experiment!" And that was how the first experiment was done, just a few doors away from my office. So I was very excited. Every morning I would come to the experiment to see how it is developing, and giving them ideas. One day Chambers took me aside and said, "Look, if you want this experiment ever to be finished, please continue with your theoretical work, and let us do our experiment without your intervention." I decided to accept his kind suggestion, and did not intervene any more. And indeed the experiment was finished,[2] but, as Professor Yang said, it was not an ideal experiment because the magnetic whisker, even though it was fairly straight, contained also lines of magnetic field that did go out. Therefore it could not be claimed that the electron beam is completely free of a magnetic field. Then, there was a better attempt[3] by Möllenstedt, who actually built a very, very thin solenoid, and did the experiment with it. But still, because it was not completely shielded, people did not accept this as a true experimental test.

This is where Tonomura enters the story. It was he who finally managed to beautifully do this experiment[4-6] with the magnetic field completely shielded by a superconductor surrounding it. And the electron beam did show the effect: the difference of one fluxon (half a fluxon from the point of the view of the electron) from the point of view of the two-charge carrier. And there was indeed a shift in interference fringes that showed the effect.

## 3. Wave or particle? Heisenberg's picture re-assessed

This is about the history. Now I want to tell you now some ideas connected with the effect which have developed since then. Let me start by saying that the way interference is introduced in books is misleading, because if we look at quantum interference, for example a two-slit quantum interference, we see a picture very similar to classical interference. Apparently in both classical and quantum interference, you see a wave: One part of the wave goes through one slit and the other part goes through the other, then the two beams meet and create an interference pattern. There seems to be no difference between the classical and the quantum picture, I now argue that there *is* a fundamental difference.

In classical theory, there is local information in the two slits that tells you what will happen when the two parts of the wave finally meet. But in quantum mechanics, what tells you where are maxima and minima is the relative *phase*, which goes through the one slit minus the phase in the other. But this phase cannot be observed, because quantum mechanics allows changing only the phase of the *whole* wave function by a constant, which therefore makes no difference. So the only thing that can be observed in quantum interference is the difference of the phases between the two slits. But no experiment could be done to say what the phase in each slit is individually. This is a truly *non-local* phenomenon, not obvious in Schrödinger's picture. Classically, when you speak about electromagnetic waves, water waves, or any classical wave giving rise to interference, there will be always a possibility of doing a local experiment at each slit that tells what the phase is. For example, if it's an electromagnetic field, the local experiments can tell what the electric field is at a given point. So the relative phase is really a difference between truly local properties in the first slit and truly local properties in the second one. But in quantum mechanics this is not true because quantum interference is truly non-local. And this non-local phenomenon remains hidden, when you think about it, in the Schrödinger representation.

So when I thought about it, I thought that perhaps one could gain a better understanding of what happens in quantum interference if one thinks about it in the Heisenberg picture instead of the Schrödinger picture. The question is as follows: what are the properties of interference that we see in the Heisenberg picture? In this picture the wave function is a constant thing: You only look at the time dependence of operators that are functions of position and momentum. How does this description catch these non-local features of quantum interference?

The insight that drove me to do it in the Heisenberg picture came from a solution of the following paradox, which I myself invented. Consider a grating with distance $L$ between the slits (Fig. 1). I send through it a very weak electron beam, one electron after the other. What will happen? We all know that the beam can go in one out of some quantized directions, so that the difference between the lengths of neighboring rays is an integer times the wavelength:

$$P_y = \sin\theta_a P = n\lambda P/L = nh/L,$$

hence only this direction will get constructive interference.

Now suppose I put solenoids behind each part of the grating, so that each solenoid, being half a fluxon, gives a relative phase of π between adjacent beams. This means that now, the new directions of constructive interference will be shifted, as shown in Fig. 1b:

$$P_y = \sin\theta_b P = (n+1/2)h/L.$$

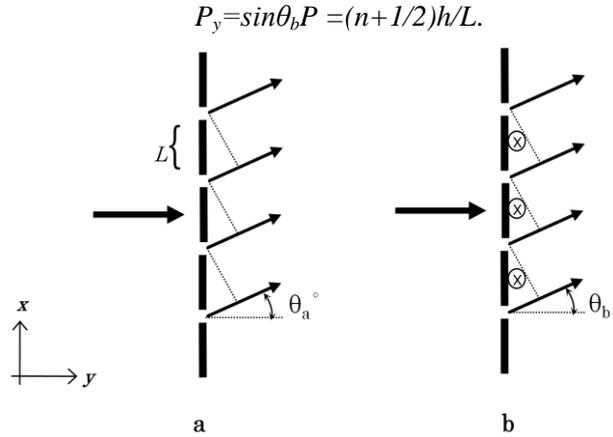

Fig. 1. Electron beam diffracting through a grating

Now, you need an extra *L/2* in order to get constructive interference. Let us see what this means in terms of momentum. If I call the vertical direction *y*, and the horizontal direction *x*; then without the solenoid, I know that the change in the electron's momentum was an integer times *h/L*. *With* the solenoids, this momentum change will be an integer plus *h/2L*. What happens to momentum conservation? I know that the sum of the momenta of the electron plus the grating plus the solenoid must be conserved. I know that the grating can give the electrons only an integer times *h/L* momentum, due to its periodicity. So it is the solenoids that must give the missing momentum, which is at least ±*h/2L*. And that happens *without any force*. I can make sure that the electron does not touch the solenoid. I can even put a barrier to make sure that the electron never touches the magnetic field. So without forces at all, we know that there is a change of momentum between the array of solenoids and the electron which is at least ±*h/2L*.

So far it's no big deal, because we know that the positions of the solenoids cannot be better defined than *L* because they have to be hidden from the electron. So the uncertainty in the momentum of the solenoid is *h/L*, which is bigger than the momentum exchange with the electron *h/2L*. But suppose now we send one electron after the other and ask what will happen after *N* electrons went by. We expect each electron to give at least ±*h/2L*. After *N* electrons go through, the array of solenoid should get something like $h\sqrt{N}/2L$. Because there is no memory, each time is like a random walk. Each time the solenoid should get either *h/2L* or -*h/2L*. So eventually, without any force, the array of solenoids should start to move. But that's against the classical intuition, because classically if the solenoids don't feel any force, they shouldn't move at all. So here was a paradox. How could you solve it? The solution led me to this new way of thinking about interference in the Heisenberg picture.

First I told myself the following. If instead of the random walk on a *straight line*, I will have a random walk on something like a *circle*, then the momentum will not keep on going but will stay bounded. And indeed once I thought about it like this, I understood the following: With no solenoid, I know that the change in the momentum is equal to some integer times *h/L*. With the solenoid, the change in the momentum is equal to an integer ± *h/2L*. That means that the relevant quantity that is changing here is not the momentum itself, but the momentum *modulo h/2L*.

I realized that this is a very important quantity to describe this kind of *topological* effects. In topological effects, you don't change just the ordinary moments of the momentum, but instead you have an exchange between the solenoids and the electrons of this modulo variable which has no classical limit. The reason is that in the classical limit, when I keep *L* fixed and *h* goes to 0, the modular momentum becomes unobservable – it is oscillating infinitely fast. This is why I began thinking that the correct way to look at quantum interference is by using this kind of *modulo* variables. Let me show you how this works.

First, let us look at what happens when we have just two slits. Let's look at the wave function of the electron just after it passes the two slits. It is the superposition of two wave packets, $\psi_1$ and $\psi_2$ (Fig. 2). The distance between them is *L*. And for simplicity I assume that there is no overlap between them, just to make the mathematics simpler.

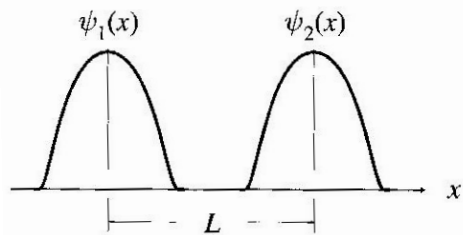

Fig. 2 Two non-overlapping wave functions on the x axis

Now I ask: What kind of physical variable will be sensitive to the relative phase between the two? Let us define

$$\psi_\alpha \equiv \frac{1}{\sqrt{2}}(\psi_1 + e^{i\alpha}\psi_2),$$

where I assume that $\psi_1$ and $\psi_2$ are just shifted by $L$ relative to each other. I now ask: Which operator, in the Heisenberg picture (being a function of position and momentum), will be sensitive to this relative phase α?

It turns out that if I look at any operator

$$f(x,p) = \sum_n \sum_m a_{mn} x^n p^m$$

symmetrized, it is insensitive to *α*. In other words, the average of this operator will be independent of *α*. Just to see why this is the case, look, for example, at the following integral:

$$\int (\psi_1^* + e^{-i\alpha}\psi_2^*) p^n (\psi_1 + e^{i\alpha}\psi_2) dx .$$

It will be the average of the *n-th* power of the momentum. You can see immediately, because the momentum acts like a derivative, that the only thing that will depend on α will be these mixtures. But because the *n-th* derivative of $\psi_1$ is only localized in one slit, its product with $\psi_2$ everywhere is 0. Therefore, there will be no dependence on α for the average of any power of the momentum. Certainly there will be no dependence on α if ever there is a power of position or the product of position and momentum.

So what is going on? Which operator is sensitive to it? Well it turns out that if I look at the operator $e^{ipL/\hbar}$, it acts like a translational operator. In order to prove that you don't necessarily have to use power expansion. You can use directly the Fourier transform, and as long as you have the function of $x$ of the Fourier transform you can show that $e^{ipL/\hbar}$ is a translational operator. So when I take $e^{ipL/\hbar}$ and calculate each average, it will be equal to $e^{i\alpha/2}$. You can easily check that, because $e^{ipL/\hbar}$ now takes $\psi_1$ and brings it to $\psi_2$, or vice-versa. You can immediately say: "but you can expand $e^{ipL/\hbar}$ as an infinite power series of momentum!" Well, when you do expand it, it looks as if it is a Taylor expansion; that is what you take from the two slits. But because there is no overlap, this function *cannot* have a Taylor expansion. The Taylor expansion doesn't converge. And therefore, the average of the power series is not equal to the series of the averages. This is an example where the average $(dp/dt)e^{ipL/\hbar}$ does indeed depend on the electric phase. But if you look separately at each one of the components in the expansion, you see that this sum doesn't converge, because there is no Taylor expansion.

I argue that this variable, which is the modulo momentum, is the correct variable to think of when considering what happens in quantum interference. It has all the right properties. In the classical image, it disappears because it doesn't exist. In the quantum limit it is indeed sensitive to the relative phase.

So the idea of thinking about quantum interference in the Heisenberg picture is the following. When a particle comes through a pair of slits, with a solenoid behind them, the solenoid will not change any of the moduli, any of the powers of the momentum, because it changes only the *relative* phase. All the momentum powers, the average of all the moments, will be unaffected by it. The only things that will be affected in the A-B effect will be the modulo variable. And therefore we can think about the A-B effect as an exchange; a non-local exchange of this modulo momentum. The modulo momentum has been conserved too. Let's look at two interacting systems and write, for example, $\cos(p_1 L/\hbar)$ as the modulo variable of the first one and $\cos(p_2 L/\hbar)$ over the second. We know that $p_1+p_2$ is a conserved quantity. Then it is very easy to show that their conservation comes from the fact they move on an ellipse. That means that if I change the modulo variable of one system, there must be also a corresponding change of the other. So I can think about the interaction between the solenoid and the wave packets of the electron as an exchange, not of momentum, but an exchange of *modulo momentum*.

This exchange happens non-locally. That means that if I look at the Heisenberg equation of motion, it is fundamentally different from the classical equation of motion. Usually people think that the fundamental difference, or the most important difference, between classical and quantum mechanics is in the kinematics; namely, that the quantum particle is described by a wave, whereas a classical particle is described by point, position, velocity. But when you look at the equation of motion, you've changed the Poisson brackets to commutators. So it looks as if the difference between the classical and quantum accounts is just corrections of the order of *h*. It turns out that this is not true. The fundamental difference between the classical and quantum equation of motion emerges when you look at the variables that are relevant for interference. Assume we have a Hamiltonian for a particle which is the free part of the Hamiltonian plus, say, some potential *V(x)*. In classical physics, it is true that if you take any function of momentum *f(p)*, the time derivative of this function is equal simply to

$$\frac{df(p)}{dt} = \frac{\partial f}{\partial p}\frac{dp}{dt} = -\frac{\partial f}{\partial p}\frac{dV}{dx},$$

meaning that in classical physics, for the particle's momentum to change, it must be in a place where there is a force. But in quantum mechanics, we have seen that the relevant quantity is the modulo variable. If we look at the time derivative, for example, of $e^{ipL/\hbar}$, we find that it is equal to

$$\frac{d}{dt}e^{ipL/\hbar} = \frac{i}{\hbar}[V(x)-V(x+L)]e^{ipL/\hbar}.$$

So you suddenly realize that in quantum mechanics, when you look at the relevant variables for interference, they include not only the potential where the particle is, but also the potential distance *L* away from the particle.

Here is a new way to think about the mystery of the two-slit interference. Feynman[7] has once made a famous statement about the double-slit experiment that implies that nobody really understands what goes on in quantum interference. I beg to differ! There is an intuitive way to understand what is going on as described in the following: You send the particle and indeed, you do not know whether it goes this way or that way. But you would like to say that it goes through one of them, only you don't know which. The minute you say that, you have to ask: "Ah, but how does the particle that goes *there* knows that the potential *here*, whether this slit is open or not?" The answer I propose is that there is a non-local equation of motion of the relevant variable, which is the modulo momentum. It tells you that if one slit is closed, it affects the modulo momentum differently than if it is open. But then you say, "Hey, wait a minute. Doesn't it break causality? If the electron is now moving *here*, and it knows whether the other slit *there* is open or not, doesn't it violate causality? After all, opening or closing the slit is a last-minute decision!" Well, quantum mechanics saves us in a very interesting way. It shows that this modulo momentum, *p mod (h/L)*, has a very interesting property: if the particle is confined to a location limited by *L*, and if you look at the modulo momentum of any *L* equal to that of the latter, that is, *p mod (h/L)* becomes completely uncertain. What does it mean "completely uncertain"? You can think about it as a circle because it is limited. If you have an equal probability for any value, then of course if you rotate the circle you will not see any difference. It turns out that when the electron goes through one slit, its modulo momentum is completely uncertain. Therefore, although the modulo momentum has an effect, it is unobservable. But when the electron's position is in superposition of being in one slit and in the other, then this modulo variable is known. And then the non-local effect can be observed.

So, the basic idea is to point out the truly non-local side of quantum interference. Nonlocality, which will later emerge even more dramatically in the EPR experiment,[8] is presented by variables that have no classical analog. These variables can be associated with each individual particle, not like the Schrödinger wave, which, in essence, is a property of an *ensemble*. The beautiful thing about the Heisenberg representation is that variables can be observed on *individual* particles. This, by the way, is what we accomplish nowadays with weak measurements.[9] So we have to learn again to think not only in the Schrödinger representation, which is very simple mathematically though very confusing intuitively. We have to learn how to observe quantum phenomena from both the Schrödinger and Heisenberg points of view.

Finally please allow me to share with you an interesting historical anecdote. I met Heisenberg in the Max Planck Institute a few years before he died. That was in the early 1960s. I met him in his office and asked him, "Professor Heisenberg, how do you think about interference phenomena using your representation?" Lo and behold, he said, "I don't know how to do it. I've never thought about it." Really! Then I offered him this idea. He was so excited and so happy about it, and I want to propose the reason for that. There is something very interesting in the development of quantum mechanics. Schrödinger and Heisenberg came practically the same time, Heisenberg a little earlier. Everybody tried to solve everything using Heisenberg's matrix representation. It was very complicated. It could solve only very few problems in the beginning. Then came Schrödinger and immediately everybody jumped on his wagon because it was so much easier to work with. Heisenberg was naturally unhappy. He tried to fight this change but could not slow that bandwagon. So the idea I proposed on that day showed him that it *is* possible to understand quantum interference in his language. He was so happy about it that I heard indirectly from his assistant, Hans-Peter Dürr, that since that day every visitor who came to visit Heisenberg, the first thing he would show them is how to understand the interference pattern in the Heisenberg language. That made him extremely happy.

Closing the circle, I think this is also the correct way to think about the A-B effect. It is an exchange of these dynamical variables.

So I recommend to everybody to pay more attention, not only to Schrödinger's representation, but also to Heisenberg's. Because if you learn to do it you will look at the Schrödinger representation as only a mathematical simplification aiding the solution of complicated problems. But then things that appear have only to do with an ensemble, *not* with individual particles. You can't say any more that there is a wave function moving in space. There is nothing like that! But if you think about Heisenberg representation, you can think about variables that can really be associated with individual particles. When you look at functions of position and momentum – these are observables. You can look at individual particles and play around with the Heisenberg equation of motion. You get much more intuition. Much of my later work on weak measurements[9,10] presents this intuition's fruits.

## 4. Discussion (Questions and answers)

Q: I was wondering whether these ideas about interference can also shed some light about collapse, which is basically a loss of non-locality. Have you thought about this? I'm sure you have.

A: Yes and No. First of all, even in the Heisenberg representation, there is no indication for collapse. The collapse will have to come from someplace else. So far, there is no way that you can say otherwise, apart from some kinds of very ugly attempts to introduce non-localities into the Schrödinger equation. Up to now there is no solution of the collapse, although I have some ideas which are not connected with this. They are connected with what will be discussed by my colleague Jeff Tollaksen. But it is not finished yet so I'm not going to talk about it. The collapse is still an open issue. But at least out of collapse I am saying there is a way to develop more intuition about what happens in the quantum domain provided you learn to think about it in the classical language, and see the difference between the classical language when you change from the dynamical equation of motion by Poisson brackets to the dynamical equation of motion by commutators. That's where the basic new difference between quantum and classical mechanics is - The non-locality of the quantum equation of motion. That is the thing that was missed.

Q: How do I think about having detectors at both slits in this language?

A: The point is that once you put the detector, in any one of them, it makes the modulo momentum completely uncertain. And that is the beautiful thing. This complete uncertainty saves you from breaking causality,

I want to emphasize the difference between complete uncertainty and the usual Heisenberg uncertainty. Heisenberg has certain principles that are quantitative. The complete uncertainty is a qualitative one. There must be some reason for it. And my question is "Why does God play dice?" If God wants to have non-local equations of motion, and at the same time to save causality, you had better have uncertainties. But if you have uncertainties which are complete, you could not detect the violation of causality. And this complete uncertainty of the modular momentum…. It's a very strong one. From it you can deduce the usual Heisenberg uncertainty, but not vice versa. So this is the more fundamental way to think about the quantum uncertainties. There is a reason why there must be uncertainties if you want to combine non-locality and causality together.

## Acknowledgements

This work has been supported in part by the Israel Science Foundation Grant No. 1125/10.